\documentclass[showpacs,preprintnumbers,amsmath,amssymb]{revtex4}

\usepackage{graphicx}
\usepackage{dcolumn}
\usepackage{bm}

\begin{document}
\def\blambda{{\hbox{\boldmath $\lambda$}}}
\def\eeta{{\hbox{\boldmath $\eta$}}}
\def\bxi{{\hbox{\boldmath $\xi$}}}

\title{Total Hadron-Hadron Cross Sections at High Energies}

\author{Wei-Ning Zhang$^{1,2}$}
\author{Cheuk-Yin Wong$^1$}

\affiliation{
$^1$Physics Division, Oak Ridge National Laboratory, Oak Ridge, TN
37831, U.S.A.\\
$^2$Department of Physics, Harbin Institute of Technology, 
Harbin, 150006, P. R. China}

\date{\today}

\begin{abstract}
We calculate total hadron-hadron cross sections at high energies using
the Low-Nussinov two-gluon model of the Pomeron.  The gluon exchange
is represented by a phenomenological potential including screened
color-Coulomb, screened confining, and spin-spin interactions. We use
bound-state wave functions obtained from a potential model for mesons,
and a Gaussian wave function for a proton.  We evaluate total cross
sections for collisions involving $\pi$, $K$, $\rho$, $\phi$, $D$,
$J/\psi$, $\psi'$, $\Upsilon$, $\Upsilon'$, and the proton.  We find
that the total cross sections increase with the square of the sum of
the root-mean square radii of the colliding hadrons, but there are
variations arising from the spin-spin interaction.  We also calculate
the total cross sections of a mixed-color charmonium state on a pion
and on a proton.  The dependence of the total cross section on the
size and the color state of the charmonium is investigated.
\end{abstract}


\maketitle

\section{Introduction}

Hadron-hadron total cross sections are important characteristics of
the dynamics of hadronic systems.  Theoretical descriptions of the
reaction process, however, remain incomplete as the mechanism of how
hadrons interact contains important aspects of non-perturbative QCD.
A phenomenological description of Donnachie and Landshoff \cite{Don92}
gives the total cross section in terms of the exchange of a Pomeron
and a Reggeon. The exchange of a Pomeron dominates the cross section
at high energies.  It corresponds to the exchange of a tower of
fictitious particles on a Regge trajectory with a Regge intercept of
approximately 1.  The exchange of a Reggeon dominates the cross
section at low energies.  It corresponds to the exchange of
$\rho,\omega,f_2,a_2...$ on the Regge trajectory with a Regge
intercept of approximately 0.5.

It is desirable to study microscopic models of total hadron-hadron
cross sections at high energies. The total cross sections provide
useful information on the sizes of hadrons and the dynamics between
them.  Furthermore, they are important input data for the
investigation of other reaction processes.  For example, in
high-energy heavy-ion collisions, the suppression of $J/\psi$
production has been proposed as a signature for the production of the
quark-gluon plasma \cite{Mat86}.  However, the suppression of $J/\psi$
production can also arise from the collision of $J/\psi$ with hadrons
at high energies. It is necessary to understand the absorption of
$J/\psi$ by hadrons before one can unambiguously identify the
quark-gluon plasma as the source of $J/\psi$ suppression
\cite{Ger88,Won96,Won98a,Kha96,Bla96,Cap97,Cas97,Vog98,Nar98,Zha00,Bla00,
Mar95,Won00a,Won00b,Bar00,Won01,Bar03,Xu03}.  While the systematic
studies of Donnachie and Landshoff provide very valuable information
on the total cross section of many hadron-hadron systems, there are
reactions (involving, for example, $J/\psi$) for which data are not
available.  They can only be evaluated theoretically
\cite{Bar92,Kha94,Mat98,Hag00,Lin00,Won2002,Won2000}.  It is therefore
useful to develop a microscopic theory of hadron-hadron collisions at
high energies.

As the Pomeron exchange is the dominant process in high-energy
hadron-hadron reactions, it is useful to model the exchange of a
Pomeron explicitly in microscopic terms.  Previously, the Pomeron
exchange in a hadron-hadron reaction was studied in terms of the
exchange of two gluons.  The two-gluon model of the Pomeron (TGMP),
first proposed by Low \cite{Low75} and Nussinov \cite{Nus75}, was
further investigated by Gunion and Soper \cite{Gun77} who introduced
an effective gluon mass to mimic the confinement of the colored gluon.
They examined the effects of the size of the bound-state hadrons and
the number of quarks they contain.  Landshoff and Nachtmann
generalized the concept of the gluon condensate to the correlation
length of the gluon field in the vacuum and discussed the short-range
nature of the effective gluon exchange between quarks in hadron-hadron
scattering \cite{Lan87}.  The dependence of the cross sections on the
sizes and the color states of the colliding hadrons was considered by
Dolej\v si and H\" ufner \cite{Dol92}.  They obtained useful
analytical and numerical results for the total cross sections of many
hadron-hadron systems.  The effects of channel coupling for the
propagation of a radially-excited hadron \cite{Huf96} was investigated
in the two-gluon model of the Pomeron by Wong \cite{Woncw96} who also
evaluated many meson-nucleon total cross sections.

We shall follow the approach of Gunion, Soper, Landshoff, Nachtmann,
Dolej\v si, H\"ufner, and Wong to study the hadron-hadron total cross
section, with the following refinements.  We shall use a set of meson
wave functions obtained from meson mass calculations \cite{Won01} to
replace the simple form factors used in earlier studies. This allows a
more detailed and systematic study of the hadron-hadron scattering
processes involving a much large set of hadrons than those
investigated previously.  Gunion and Soper \cite{Gun77}, Dolej\v si
and H\"ufner \cite{Dol92}, and Wong \cite{Woncw96} made use of only a
single screened color-Coulomb potential to represent the
non-perturbative gluon exchange.  We shall instead represent the gluon
exchange between the constituents of one hadron and constituents of
the other hadron by a phenomenological potential containing screened
color-Coulomb, screened color-confining, and spin-spin interactions
\cite{Kar88, Won99,Won01,Won02}.  Screening arises when the
interaction between a constituent of one hadron and a constituent of
another hadron occurs at large distances for which the production of
dynamical quarks screens the interaction.  The use of a more
complicated potential allows a greater degree of flexibility, which
provides a better description of non-perturbative effects in
scattering.  It also enables one to use reasonable strong interaction
coupling constants, as the constraint of using a single screened
color-Coulomb interaction may sometimes lead to values of $\alpha_s$
larger than unity \cite{Dol92}.  In addition, the spin-spin
interaction is known to be important in non-relativistic quark models.
Its phenomenological incorporation here will allow us to evaluate the
variation of the total cross section in reactions of mesons with
different total spin quantum numbers and quarks of different flavors.

Similar to earlier works of Dolej\v si and H\"ufner \cite{Dol92}, and
Wong \cite{Woncw96}, we shall first obtain the the screening mass and
the strength of the interaction to reproduce the experimental $pp$,
$\pi p$, and $Kp$ cross sections at $\sqrt{s}=20$ GeV.  We then use
the theoretical meson wave functions to calculate other cross
sections.  Based on the energy dependence of the phenomenological
description of total hadron-hadron cross sections found by Donnachie
and Landshoff \cite{Don92}, we shall then extend our results to higher
energies.

In reactions involving heavy quarkonium production in nuclear
collisions, the initial heavy quark pair produced as a result of a
parton-parton collision is a coherent state of a color admixture and
different projections of this state will lead to different final heavy
quarkonia \cite{Bod95,Won96,Won97,Won98,Won99a,Huf01,Huf00}.  We will also
examine the color dependence of the cross section, in order to
understand how a colored heavy quark pair propagates in a hadronic
medium \cite{Dol92,Won96,Won97,Won98,Won99a,Huf01,Huf00}.  The nucleons with
which the produced charmonium collide can also acquire a color as a
result of prior collisions and may collide with a color-singlet
charmonium \cite{Huf01,Huf00}.  In this paper, we shall limit our
discussion to the problem of the interaction of a colored charmonium
state with a color-singlet hadron.

Our approach to study the total cross section in the TGMP differs from
that of the additive quark model \cite{Lev65} and the dipole model of
photoproduction \cite{Huf00a}.  The total cross sections obtained in
different models can naturally be different.  Careful comparison of
these different results with experimental data will be useful to
determine the importance of various scattering mechanisms that are
present in the interaction of hadrons.

This paper is organized as follows.  In Sec. II we describe the model
used to calculate the hadron-hadron elastic scattering amplitude at
high energies.  In Sec. III we evaluate the spin matrix element for
meson-meson, meson-proton, and proton-proton scattering.  The
calculation of the spatial matrix element is discussed in Sec. IV.  In
Sec. V we present results for total hadron-hadron cross sections and
the energy dependence of the total hadron-hadron cross sections.  In
Sec. VI we evaluate the total cross sections of a mixed-color
charmonium state scattering on a pion or a proton.  The dependences of
the total cross sections on the size and the color of the initial
charmonium state are investigated.  Finally, we present our summary
and conclusions in Sec. VII.

\section{Two-Gluon Model of the Pomeron}

To calculate the total cross section in a hadron-hadron collision, we
consider the hadron-hadron elastic scattering amplitude $\cal A$.
According to the optical theorem, the imaginary part of the elastic
scattering amplitude at zero momentum transfer gives the total
hadron-hadron cross section.

The elastic scattering process between two hadrons at high energies is
dominated by the exchange of a Pomeron, as evidenced by the slow
variation of the total cross section with energy, the property of
having an approximately purely imaginary forward scattering amplitude,
and the dominance of no quantum number flow in the fragmentation
region in forward inelastic processes.  The phenomenological treatment
of the total cross section as arising from the exchange of a Pomeron
with a Regge trajectory intercept of 1.0808, and the exchange of a
Reggeon from the trajectory of $\rho,\omega,f_2,a_2...$ with an
intercept of 0.5475, gives an excellent representation of many
hadron-hadron total cross sections \cite{Don92}.  The small deviations
of these Regge trajectory intercepts from the lowest-order
expectations of 1.0 and 0.5 respectively have been attributed to
higher-order effects \cite{Don92}.

In this work, we shall model the exchange of a Pomeron in terms of the
exchange of gluons.  As the exchange of a single gluon would lead to
an exchange of color, the lowest number of gluon exchanges with no net
color flow is the exchange of two gluons, proposed first by Low
\cite{Low75} and Nussinov \cite{Nus75} and studied by Gunion, Soper,
Landshoff, Nachtmann, Dolej\v si, H\"ufner, and Wong
\cite{Gun77,Lan87,Dol92,Woncw96}.

Accordingly, we express the total hadron-hadron cross section in terms
of the hadron-hadron elastic scattering amplitude at zero
four-momentum transfer squared $t$
\begin{equation}
\label{optical}
\sigma_{\rm tot} = \frac{ 1}{s} \,{\rm Im} \,{\cal A} (s,t=0) \, .
\end{equation}

Following Gunion and Soper \cite{Gun77} and Dolej\v si and H\"ufner
\cite{Dol92}, we obtain the forward scattering amplitude $\cal A$ by
including diagrams of the type shown in Fig.\ 1.  As the elastic
scattering process does not change the hadron internal structure, the
initial and final hadron states are identical.  In Fig.\ 1,
constituents $i$ or $j$ in hadron I interact with constituents $k$ and
$l$ in hadron II by exchanging two gluons.  As the elastic process
involves important aspects of non-perturbative QCD, it is reasonable
to represent the corresponding gluon exchange between constituents $i$
and $k$ in terms of a phenomenological potential $V(ik)$
possessing non-perturbative elements of color-Coulomb, confinement,
and spin-spin interactions.  The complete amplitude is obtained by
summing over all possible diagrams representing the exchange of two
gluons between all constituents $\{i,j\}$ of hadron I with all
constituents $\{k,l\}$ of hadron II \cite{Dol92}:
\begin{eqnarray}
\label{ampl}
{\cal A}(s,t) = \frac{ i s }{ 16 } \sum_{i,j,k,l} &\Bigg\{ &\int
{\rm d}^2 {\bf b} ~ e^{-i {\bf Q} \cdot {\bf b} } \bigg[\langle\,
C_{\rm I}\,C_{\rm II}\,\chi_{_{\rm I}}\,\chi_{_{\rm II}}\,\Psi_{\rm I}
\,\Psi_{\rm II}\,|\,V_{\rm T}(ik)\,\nonumber\\
& & \times V_{\rm T}(jl)\,|\,C_{\rm I}\,
C_{\rm II}\,\chi_{_{\rm I}}\,\chi_{_{\rm I}}\,\Psi_{\rm I}\,
\Psi_{\rm II}\,\rangle \bigg]\Bigg\}\,,
\end{eqnarray}
where ${\bf b}$ is the impact parameter, ${\bf Q}^2=-t$, $C_{\rm I}$
and $C_{\rm II}$ are the color wave functions, $\chi_{_{\rm I}}$ and
$\chi_{_{\rm II}}$ are the spin wave functions, and $~\Psi_{\rm I}$
and $\Psi_{\rm II}$ are the spatial wave functions of hadron I and
hadron II, respectively.  The square brackets represent the averaging
over the color, spin, and internal spatial degrees of freedom of
hadrons I and II.  The sum $\sum_{i,j,k,l}$ is over all two-gluon
exchange diagrams.  

At high energies when the average intrinsic quark momentum inside a
hadron is small compared to the center-of-mass energy, the case of a
small momentum transfer in a two-gluon exchange diagram can be
approximated as a loop diagram in which the quark lines on the loop
can be treated as on the mass shell and represented by two delta
functions (as in Eq.\ (A.11) of Ref. \cite{Lan87}).  After integrating
over the time-like and the longitudinal loop momenta using these two
delta functions, the resultant effective gluon propagators involve
only transverse momenta (or equivalently only transverse coordinates)
\cite{Gun77,Lan87,Dol92}.  When we represent the effective gluon
propagator phenomenologically by an interaction, the interaction
$V_{\rm T} (ik)$ in transverse coordinates, written explicitly as
$V_{\rm T}({\bf r}_{\rm T})$, can be obtained by integrating the
interaction $V ({\bf r})$ over the longitudinal coordinate $z$,
\begin{eqnarray}
\label{eq3}
V_{\rm T}(ik)=V_{\rm T}({\bf r}_{\rm T}) = \int_{-\infty}^{\infty}
dz~V({\bf r}),
\end{eqnarray}
where ${\bf r}$ is the three-dimensional relative coordinate between
the interacting particles $i$ and $k$ and ${\bf r}_{\rm T}$ its
transverse component.  Defining the Fourier transform of $V({\bf r})$
as ${\tilde V}({\bf k})$ by
\begin{eqnarray}
\label{eq4}
V({\bf r})=\int \frac {d{\bf k}} {(2 \pi)^3}
~ e^{i {\bf k} \cdot {\bf r}}
~{\tilde V}({\bf k}),
\end{eqnarray}
we can obtain from Eqs.\ (\ref{eq3}) and (\ref{eq4}) 
\begin{eqnarray}
\label{ftrans2}
V_{\rm T}({\bf r}_{\rm T}) 
=\int \frac{d{\bf k}_{\rm T}}{(2 \pi)^2 }~ 
e^{i {\bf k}_{\rm T} \cdot {\bf r}_{\rm T}}
~{\tilde V}({\bf k})\biggr |_{k_z=0}.
\end{eqnarray}
Thus, ${\tilde V}_{\rm T}({\bf k}_{\rm T})$, the 2-dimensional Fourier
transform of $V_{\rm T}({\bf r}_{\rm T})$, is just the 3-dimensional
Fourier transform ${\tilde V}({\bf k})$ evaluated at $k_z=0$.

As the interaction between the constituents takes place at large
distances, we are well advised to use a screened potential to
represent the effects of dynamical quarks.  As pointed out by
Landshoff and Nachtmann \cite{Lan87}, two quarks in two hadrons can
exchange a nonperturbative gluon only if they pass within a short
distance of each other.  This short correlation length of the gluon
field is another manifestation of the screening phenomenon.  We employ
screened color-Coulomb (Yukawa) and screened confining (exponential)
potentials \cite{Kar88,Won99,Won02} and a spin-spin interaction
\cite{Won01,Won02} for our interaction $V({\bf r})$.  The interaction
between quark $i$ in hadron I and quark $k$ in hadron II in
three-dimensional relative coordinate ${\bf r}$ can be written as
\begin{eqnarray}
\label{vint}
V({\bf r}) &=& \frac{\blambda_i}{ 2}\cdot{{\blambda_k} \over 2}\,
\bigg\{ V_{\rm Yukawa}({\bf r})+V_{\rm exponential}({\bf r})+V_{\rm
spin-spin}({\bf r}) \bigg\} \nonumber\\
&=&\frac{\blambda_i}{ 2}\cdot{{\blambda_k} \over 2}\,\bigg\{ { {\alpha_s 
e^{- \mu r } } \over {r} } + { {3b_0} \over {4\mu} } 
e^{- \mu r } - { {8 \pi \alpha_s} \over {3 m_i m_k} } ({\bf s}_i 
\cdot {\bf s}_k ) \bigg( { d^3 \over \pi^{3/2} } \bigg) e^{-d^2 r^2} 
\bigg\} \, ,
\end{eqnarray}
where $\blambda_{\,i}$ is the Gell-Mann matrix, and $\blambda_{\,i}$ is 
replaced by $-\blambda_{\,i}^{\rm T}$ for an antiquark $i$.  The
quantities $\alpha_s$, $\mu$, $b_0$, and $d$ are the strong coupling
constant, the effective screening parameter, the effective
string-tension, and the spin-spin potential width
parameter.  The Fourier transform of the spatial and spin part of the
potential is
\begin{eqnarray}
\label{vftrans3}
{\widetilde V}({\bf k})=\bigg[{{4\pi \alpha_s}\over{ {\bf k}^2 + \mu^2 }}
+ { {6\pi b_0} \over { ( {\bf k}^2 + \mu^2 )^2 } }
- { {8\pi \alpha_s} \over {3 m_i m_k} } {\bf s}_i \cdot {\bf s}_k 
\,e^{-{\bf k}^2 /4 d^2} \bigg] \,.
\end{eqnarray}
We have used an exponential interaction in Eq.\ (\ref{vint}) to
represent the screened confining interaction as it contains the
appropriate properties.  At short distances for which screening effect
is small, the exponential interaction gives the linear interaction
$b_0 r$, and at large distances for which the confining interaction is
screened by the production of dynamical quarks, it approaches zero.
In terms of the Fourier transform, screening is represented by a
non-vanishing value of the screening mass $\mu$ is Eq.\
(\ref{vftrans3}).  The use of a screened confining interaction is
necessary here.  If we assume an unscreened linear-confining
interaction, corresponding to taking $\mu=0$ in the $6\pi b_0/({\bf
k}_{\rm T}+\mu^2)^2$ term in Eq.\ (\ref{vftrans3})), we shall see in Section
V that the total hadron-hadron cross section will be singular.

The square brackets in Eq. (\ref{ampl}) can be written as
\begin{eqnarray}
\label{cssp}
& & \bigg [ \langle \, C_{\rm I}\, C_{\rm II}\,\chi_{_{\rm I}}\,
\chi_{_{\rm II}}\,\Psi_{\rm I}\,\Psi_{\rm II}\,|\,V_{\rm T}(ik)\,
V_{\rm T}(jl)\,|\,C_{\rm I}\,C_{\rm II}\,\chi_{_{\rm I}}\,\chi_{_{\rm 
I}}\,\Psi_{\rm I}\,\Psi_{\rm II}\,\rangle \bigg ] \nonumber\\
& & = \bigg \{ C_{ijkl} \sum_{a,b=1}^8 \langle\,C_{\rm I}\,|\, \lambda^a_i 
\lambda^b_j \, |\,C_{\rm I}\,\rangle \langle\,C_{\rm II}\,|\,\lambda^a_k 
\lambda^b_l\,|\,C_{\rm II}\,\rangle \bigg \} \bigg \{ {1\over N_{\chi}} 
\sum_{\chi_{_{\rm I, II}}}
\nonumber\\
& & \times \langle\,\chi_{_{\rm I}}\,\chi_{_{\rm II}}\,\Psi_{\rm I}({\bf 
r}_{\rm I})\,\Psi_{\rm II}({\bf r}_{\rm II})\,|\,V_{\rm T}({\bf r}_{ik{\rm T}})
V_{\rm T}({\bf r}_{jl{\rm T}})\,|\,\chi_{_{\rm I}}\,\chi_{_{\rm II}}\,
\Psi_{\rm I}({\bf r}_{\rm I})\,
\Psi_{\rm II}({\bf r}_{\rm II})\,\rangle \bigg\}\,,
\end{eqnarray}
where the first pair of curly brackets is the sum of color matrix
elements, $C_{ijkl}$ is the product of the four factors $C_i$, $C_j$,
$C_k$, and $C_l$, each of which is 1 for a quark and $-1$ for an
antiquark, and $a$ and $b$ are gluon color component labels.  In the
second pair of curly brackets, the sum is over all spin states of
hadrons I and II (total number $N_{\chi}$).

The color wave function for a color-singlet meson is
\begin{equation}
\label{cssm}
|~C^1\,\rangle ={1 \over {\sqrt{3}}} \sum_{\alpha =1}^3 |~\alpha \,
{\bar \alpha} \,\rangle\, ,
\end{equation}
where $\alpha$ ($\alpha=1, 2, 3$) is the quark color label and ${\bar
\alpha}$ is the corresponding antiquark color label.  The color matrix
element is
\begin{equation}
\label{csmm}
\langle\, C^1~| \lambda^a_m \lambda^b_n |~C^1\,\rangle = {2 \over 3} ~
\delta^{ab} \, .
\end{equation}
For a color-singlet nucleon we have 
\begin{equation}
\label{cssn}
|~C^1\,\rangle = {1 \over {\sqrt{6}}} \sum_{\alpha, \beta, \gamma =1}^3 
\varepsilon_{\alpha \beta \gamma} ~ |~\alpha ~\beta~\gamma \,\rangle \, ,
\end{equation}
and
\begin{eqnarray}
\label{csmn}
\langle\, C^1~| \lambda^a_m \lambda^b_n |~C^1 \,\rangle =  
\begin{cases}
{2 \over 3} ~\delta^{ab}, &  ~~~~{\rm for}~~m = n , \\
-{1 \over 3} ~\delta^{ab}, & ~~~~{\rm for}~~m \ne n \,. 
\end{cases} 
\end{eqnarray} 

\section{Spin matrix elements}

From Eqs. (\ref{ampl}) and (\ref{vint}), the matrix element for the
spatial and spin part can be separated into a sum of products of spin
matrix elements and spatial matrix elements. The first-order spin-spin
matrix elements and the second-order spin-spin matrix elements are
\begin{equation}
\label{1smf}
s^{(1)}_{ik}=\langle\,\chi_{_{\rm I}}\,\chi_{_{\rm II}}~|\,({\bf s}_i \cdot 
{\bf s}_k) \,|~\chi_{_{\rm I}}\,\chi_{_{\rm II}}\,\rangle \, ,
\end{equation}
\begin{equation}
\label{2smf}
s^{(2)}_{ikjl} = \langle\,\chi_{_{\rm I}}\,\chi_{_{\rm II}}~|\,({\bf s}_i 
\cdot {\bf s}_k)({\bf s}_j \cdot {\bf s}_l) \,|~\chi_{_{\rm I}}\,
\chi_{_{\rm II}}\,\rangle \, .
\end{equation}
We shall give results for these quantities for meson-meson,
meson-proton, and proton-proton collisions separately.

\subsection{The meson$-$meson case}

Consider meson I with constituents $i$ and $j$ coupled to total spin
$S_{\rm I}$, and meson II with constituents $k$ and $l$ coupled to
total spin $S_{\rm II}$.  We would like to recouple the constituents
so that the operators $({\bf s}_i \cdot {\bf s}_k)$ and $({\bf s}_j
\cdot {\bf s}_l)$ have simple eigenvalues.  We have
\begin{eqnarray}
\label{ssm}
& &|\,\chi_{_{\rm I}}\,\chi_{_{\rm II}}\,\rangle =|\,(ij)^{S_{\rm 
I}}_{S_{\rm Iz}}\,(kl)^{S_{\rm II}}_{S_{\rm IIz}}\,\rangle = \sum_{S,S_z} 
(S S_z |S_{\rm I} S_{\rm Iz} S_{\rm II} S_{\rm IIz} )\,|\,[(ij)^{S_{\rm I}} 
(kl)^{S_{\rm II}}]^S_{S_z}\,\rangle \nonumber\\
& &=\sum_{S,S_z} \sum_{S_{ik},S_{jl}} (S S_z |S_{\rm I} S_{\rm Iz} S_{\rm 
II} S_{\rm IIz} )\,{\hat S}_{\rm I} {\hat S}_{\rm II} {\hat S}_{ik} {\hat 
S}_{jl} \left \{
\begin{matrix}
s_i & s_j & S_{\rm I} \cr
s_k & s_l & S_{\rm II}\cr
S_{ik} & S_{jl} & S \cr 
\end{matrix}
\right \}
|\,[(ik)^{S_{ik}} (jl)^{S_{jl}}]^S_{S_z}\,\rangle \nonumber\\
& &=\sum_{S,S_z} \sum_{S_{il},S_{jk}} (S S_z |S_{\rm I} S_{\rm Iz} S_{\rm
II} S_{\rm IIz} )\,(-1)^{S_{\rm II} -s_k -s_l} {\hat S}_{\rm I} {\hat S}_{\rm 
II} {\hat S}_{il} {\hat S}_{jk} \left \{
\begin{matrix}
s_i & s_j & S_{\rm I}\\
s_l & s_k & S_{\rm II}\\
S_{il} & S_{jk} & S 
\end{matrix} \right \}\,\nonumber\\
& & ~~~ \times |\,[(il)^{S_{il}} (jk)^{S_{jk}}]^S_{S_z}\,\rangle \, ,
\end{eqnarray}
where ${\hat S}=\sqrt{2S+1}~$.  From the above equation, the
matrix element $s_{ik}^{(1)}=\langle\,\chi_{_{\rm I}}\,\chi_{_{\rm II}}\,
|\,({\bf s}_i \cdot {\bf s}_k)\,|\,\chi_{_{\rm I}}\,\chi_{_{\rm II}}\,
\rangle$ is
\begin{eqnarray}
\label{1smm}
s^{(1)}_{ik} = \sum_{S,S_z} \sum_{S_{ik},S_{jl}} 
|( S S_z | S_{\rm I} S_{\rm Iz} S_{\rm II} S_{\rm IIz} )|^2\,
{\hat S}^2_{\rm I} {\hat S}^2_{\rm II} {\hat S}^2_{ik} {\hat S}^2_{jl} 
\left \{
\begin{matrix}
s_i & s_j & S_{\rm I}\\
s_k & s_l & S_{\rm II}\\
S_{ik} & S_{jl} & S
\end{matrix}
\right \}^2\,{1 \over 2}\,\big[S_{ik}(S_{ik}+1) -{3\over 2}\big] \, ,
\end{eqnarray}
and the matrix element $s_{ikjl}^{(2)}=\langle\,\chi_{_{\rm
I}}\,\chi_{_{\rm II}}\,|\, ({\bf s}_i \cdot {\bf s}_k)({\bf s}_j \cdot
{\bf s}_l) \,|\, \chi_{_{\rm I}}\,\chi_{_{\rm II}}\,\rangle$ are
\begin{eqnarray}
\label{2smm}
s_{ikik}^{(2)}
&=&\sum_{S,S_z} \sum_{S_{ik},S_{jl}} |(S S_z |S_{\rm I} S_{\rm Iz} S_{\rm 
II} S_{\rm IIz} )|^2 \,{\hat S}_{\rm I}^2 {\hat S}_{\rm II}^2 {\hat S}_{ik}^2 
{\hat S}_{jl}^2 \left \{
\begin{matrix}
s_i & s_j & S_{\rm I} \cr
s_k & s_l & S_{\rm II}\cr
S_{ik} & S_{jl} & S \cr 
\end{matrix}
\right \}^2 
\left \{ \frac{1}{2}
\big[S_{ik}(S_{ik}+1) -{3\over 2}\big]\right \}^2 ~,
\\
s^{(2)}_{ikil} &=& \sum_{S,S_z} \sum_{S_{ik},S_{jl}} 
\sum_{S_{il},S_{jk}} 
|(  S S_z | S_{\rm I} S_{\rm Iz} S_{\rm II} S_{\rm IIz} )|^2\, 
(-1)^{S_{\rm II} -s_k -s_l} 
(-1)^{S_{jl} -s_j -s_l} (-1)^{S_{jk} -s_j -s_k} \nonumber\\
& & \times {\hat S}^2_{\rm I} {\hat S}^2_{\rm II} {\hat S}^2_{ik} {\hat 
S}^2_{jl} {\hat S}^2_{il} {\hat S}^2_{jk} \left \{
\begin{matrix}
s_i & s_j & S_{\rm I}\\
s_k & s_l & S_{\rm II}\\
S_{ik} & S_{jl} & S
\end{matrix} \right\} \left\{
\begin{matrix}
s_i & s_k & S_{ik}\\
s_l & s_j & S_{jl}\\
S_{il} & S_{jk} & S
\end{matrix} \right\} \left\{
\begin{matrix}
s_i & s_j & S_{\rm I}\\
s_l & s_k & S_{\rm II}\\
S_{il} & S_{jk} & S
\end{matrix} \right\} \nonumber\\
& & \times{1 \over 2}\,\big [S_{ik}(S_{ik}+1) - {3\over 2}\big ] \,
\times {1 \over 2}\,\big [S_{il}(S_{il}+1) - {3\over 2}\big] \, ,
~~~{\rm for~} k\ne l~,
\\
s^{(2)}_{ikjl}
&=&\sum_{S,S_z} \sum_{S_{ik},S_{jl}} |(S S_z |S_{\rm I} S_{\rm Iz} S_{\rm 
II} S_{\rm IIz} )|^2 \,{\hat S}_{\rm I}^2 {\hat S}_{\rm II}^2 {\hat S}_{ik}^2 
{\hat S}_{jl}^2 \left \{
\begin{matrix}
s_i & s_j & S_{\rm I} \cr
s_k & s_l & S_{\rm II}\cr
S_{ik} & S_{jl} & S \cr 
\end{matrix}
\right \}^2 
\frac{1}{2}
\big[S_{ik}(S_{ik}+1) -{3\over 2}\big]
\nonumber \\
& & 
\times \frac{1}{2}
\big[S_{jl}(S_{jl}+1) -{3\over 2}\big],  
~~~{\rm for}~i\ne j {\rm~ and ~} k\ne l~,
\end{eqnarray}
where $i,j=1,2;\,k,l=3,4$. 

\subsection{The meson$-$proton and proton$-$proton cases}

The spin and flavor wave function for a proton in the $S_z =1/2$ state
is
\begin{eqnarray}
\label{ssp1}
|~p^{1/2}_{1/2}\,\rangle = {1 \over {\sqrt{18}}}&\big(& 2\,|\,u_+u_+d_- 
\,\rangle - \,|\,u_+u_-d_+\,\rangle - \,|\,u_-u_+d_+\,\rangle \nonumber\\
&+& 2\,|\,u_+d_-u_+\,\rangle -\,|\,u_+d_+u_-\,\rangle - \,|\,u_-d_+u_+\,
\rangle \nonumber\\
&+& 2\,|\,d_-u_+u_+\,\rangle -\,|\,d_+u_+u_-\,\rangle - \,|\,d_+d_-u_+\,
\rangle \,\big) \, .
\end{eqnarray}
Denoting the two $u$-quarks by 3 and 4, the $d$-quark by 5, and the
spinor of quark $i$ as $i^s_{s_z}$, we have
\begin{eqnarray}
\label{ssp2}
|~p^{1/2}_{1/2}\,\rangle = {1 \over {\sqrt{18}}}&\big(& 2\,\big|\,
3^{1/2}_{1/2}~4^{1/2}_{1/2} ~5^{1/2}_{-1/2}\,\rangle - \,\big|\,
3^{1/2}_{1/2} ~4^{1/2}_{-1/2} ~5^{1/2}_{1/2}\,\rangle - \,\big|\,
3^{1/2}_{-1/2} ~4^{1/2}_{1/2} ~5^{1/2}_{1/2}\,\rangle
\nonumber\\
&+& 2\,\big|\,3^{1/2}_{1/2} ~5^{1/2}_{-1/2} ~4^{1/2}_{1/2}\,\rangle - 
\,\big|\,3^{1/2}_{1/2} ~5^{1/2}_{1/2} ~4^{1/2}_{-1/2}\,\rangle - \,
\big|\,3^{1/2}_{-1/2} ~5^{1/2}_{1/2} ~4^{1/2}_{1/2}\,\rangle\nonumber\\
&+& 2\,\big|\,5^{1/2}_{-1/2} ~3^{1/2}_{1/2} ~4^{1/2}_{1/2}\,\rangle - \,
\big|\,5^{1/2}_{1/2} ~3^{1/2}_{1/2} ~4^{1/2}_{-1/2}\,\rangle - \,\big|\,
5^{1/2}_{1/2} ~3^{1/2}_{-1/2} ~4^{1/2}_{1/2}\,\rangle \,\big) \, .
\end{eqnarray}
We can couple the spin of any two of 3, 4, and 5.  For example,
coupling 3 and 4, we have
\begin{eqnarray}
\label{sspc}
|~p^{1/2}_{1/2}\,\rangle &=& {1 \over {\sqrt{18}}} \bigg[ \,2\,\bigg({1 
\over 2} {1 \over 2} {1 \over 2} {1 \over 2} \bigg| 1 1 \bigg) \bigg( 
\big|\,(34)^1_1 ~5^{1/2}_{-1/2}\,\rangle_{(1)} +\big|\,(34)^1_1 ~
5^{1/2}_{-1/2}\,\rangle_{(4)} +\big|\,(34)^1_1 ~5^{1/2}_{-1/2}\,\rangle_{(7)} 
\bigg) \nonumber\\
& & - \sum_{S_{34},S_{34z}} \bigg({1 \over 2} {1 \over 2} {1 \over 2} -
{1 \over 2} \bigg| S_{34} S_{34z} \bigg) \bigg( |\,(34)^{S_{34}}_{S_{34z}} 
\,5^{1/2}_{1/2}\,\rangle_{(2)} -\,|\,(34)^{S_{34}}_{S_{34z}}\,5^{1/2}_{1/2}\,
\rangle_{(5)} \nonumber\\ 
& &~~~~~~~~~~~~- \,|\,(34)^{S_{34}}_{S_{34z}}\,5^{1/2}_{1/2}\,\rangle_{(8)} 
\bigg) \nonumber\\
& & + \sum_{S_{34},S_{34z}} \bigg({1 \over 2} -{1 \over 2} {1 \over 2} 
{1 \over 2} \bigg| S_{34} S_{34z} \bigg) \bigg( |\,(34)^{S_{34}}_{S_{34z}}
\,5^{1/2}_{1/2}\,\rangle_{(3)} -\,|\,(34)^{S_{34}}_{S_{34z}}\,5^{1/2}_{1/2}
\,\rangle_{(6)} \nonumber\\ 
& &~~~~~~~~~~~~- \,|\,(34)^{S_{34}}_{S_{34z}}\,5^{1/2}_{1/2}\,\rangle_{(9)} 
\bigg) \bigg] \nonumber\\
&\equiv& |\,(34)\,5\,\rangle \,,
\end{eqnarray}
where the subscript $(n)$ labels the flavor component in the $n$-th
term in Eq. (\ref{ssp2}) (i.e., ${}_{(n)}\langle \,(kl)_{S_{klz}}
^{S_{kl}}l'\,|\, (kl)_{S_{klz}'}^{S_{kl}'} l'\,\rangle_{(n')}
=\delta(S_{klz}S_{klz}')
\delta(S_{kl}S_{kl}') \delta(nn')~)$.  We can also write out the other
coupling expressions of $|~p^{1/2}_{1/2}\,\rangle$, for example,
$|\,(35)\,4\, \rangle$ and $|\,(45)\,3\,\rangle$.  One can obtain a
similar decomposition for $|~p^{1/2}_{-1/2}\,\rangle$.

Thus, the first-order and second-order spin-spin matrix elements for a
meson-proton scattering can be expressed as
\begin{eqnarray}
\label{1smp}
s^{(1)}_{ik} &=& \langle\,(ij)^{S_{\rm I}}_{S_z}~[(kl)\,l'\,]\,|\,(s_i \cdot 
s_k)\,|\,(ij)^{S_{\rm I}}_{S_z}~[(kl)\,l'\,]\,\rangle \nonumber\\
&=& \langle\,(ij)^{S_{\rm I}}_{S_z}\,(kl)\,|\,(s_i \cdot s_k)\,
|\,(ij)^{S_{\rm I}}_{S_z}\,(kl)\,\rangle\,,
\end{eqnarray}
\begin{eqnarray}
\label{2smp}
s^{(2)}_{ikjl} &=& \langle\,(ij)^{S_{\rm I}}_{S_z}~[(kl)\,l'\,]\,|\,(s_i \cdot 
s_k) (s_j \cdot s_l)\,|\,(ij)^{S_{\rm I}}_{S_z}~[(kl)\,l'\,]\,\rangle 
\nonumber\\
&=& \langle\,(ij)^{S_{\rm I}}_{S_z}\,(kl)\,|\,(s_i \cdot s_k) (s_j \cdot 
s_l)\, |\,(ij)^{S_{\rm I}}_{S_z}\,(kl)\,\rangle \,,
\end{eqnarray}
where $l'$ is a spectator, $i,j=1,2$, and $k,l=3,4,5$.  Each term in
Eqs.\ (\ref{1smp}) and (\ref{2smp}) can be calculated in the same way
as in the meson-meson case.  For a proton-proton scattering, we have
\begin{eqnarray}
\label{1spp}
s^{(1)}_{ik} &=& \langle\,[(ij)\,j'\,]~[(kl)\,l'\,]\,|\,(s_i \cdot
s_k)\,|\,[(ij)\,j'\,]~[(kl)\,l'\,]\,\rangle \nonumber\\
&=& \langle\,(ij)\,(kl)\,|\,(s_i \cdot s_k)\,
|\,(ij)\,(kl)\,\rangle\,,
\end{eqnarray}
\begin{eqnarray}
\label{2spp}
s^{(2)}_{ikjl} &=& \langle\,[(ij)\,j']~[(kl)\,l'\,]\,|\,(s_i \cdot
s_k) (s_j \cdot s_l)\,|\,[(ij)\,j']~[(kl)\,l'\,]\,\rangle
\nonumber\\
&=& \langle\,(ij)\,(kl)\,|\,(s_i \cdot s_k) (s_j \cdot
s_l)\, |\,(ij)\,(kl)\,\rangle \,,
\end{eqnarray}
where $j'$ and $l'$ are spectators, $i,j=1,2,3$, and $k,l=4,5,6$.  Each
term in Eqs.\ (\ref{1spp}) and (\ref{2spp}) can be calculated similar
to the meson-meson case. 

\section{Spatial Matrix Elements}

\subsection{The meson$-$meson case}

Using a set of Gaussian basis states with different widths, Wong
$et~al.$ previously obtained bound-state meson wave functions from a
non-relativistic quark model \cite{Won01}.  The model assumes
color-Coulomb, linear-confining, and spin-spin interactions.  It gives
meson masses in reasonable agreement with experimental data.  The
meson wave functions are tabulated in Table IV of Ref.\ \cite{Won01}.
We shall use these meson wave functions in our calculations.  They are
represented by a nonorthogonal sum of Gaussian basis functions of
different widths,
\begin{eqnarray}
\label{wfm}
\Psi ({\bf r}) = \sum_{n=1}^6 a_n \phi_n ({\bf r}) 
= \sum_{n=1}^6 a_n \big({ {n \beta^2} \over \pi } \big)^{3/4} e^{-{{n \beta^2} 
\over 2} r^2} \,,
\end{eqnarray}
where ${\bf r}=({\bf r}_{\rm T},z)$ is the three-dimensional coordinate,
$a_n$ is the coefficient of the $n$-th component, and $\beta$ is a
parameter characterizing the width of the basis functions in momentum
space.

The transverse coordinates of the quarks and antiquarks in meson I and II 
can be expressed as
\begin{equation}
\label{coorm}
{\bf r}_{\rm 1T,2T} = { {\bf b} \over 2} \pm { {\bf r}_{\rm IT} \over 2} \,,
~~~~~~~~~~~~~~~~
{\bf r}_{\rm 3T,4T} = -{{\bf b} \over 2} \pm { {\bf r}_{\rm IIT} \over 2} \,,
\end{equation} 
where ${\bf r}_{\rm IT}$ or ${\bf r}_{\rm IIT}$ is the relative
transverse coordinate of the quark and antiquark in meson I or II
respectively.  The relative transverse coordinate of quark (or
antiquark) $i$ in meson I and quark (or antiquark) $k$ in meson II can
be expressed as
\begin{eqnarray}
\label{rcoorm}
& &{\bf r}_{ik \rm T}={\bf r}_{i \rm T} - {\bf r}_{k \rm T} = {\bf b} + 
{C_{{\rm I}i} \over 2} {\bf r}_{\rm IT} + {C_{{\rm II}k} \over 2} {\bf 
r}_{\rm IIT} \,, \nonumber\\[1ex]
& &( i=1,2,~~k=3,4,~~C_{{\rm I}1}=1,~~C_{{\rm I}2}=-1,~~
C_{{\rm II}3}=-1,~~C_{{\rm II}4}=1 \,.)
\end{eqnarray} 

From Eqs.   (\ref{ampl}), (\ref{ftrans2}), (\ref{cssp}), (\ref{wfm}),
and (\ref{rcoorm}), the spatial matrix element for a meson-meson
scattering is
\begin{eqnarray}
\label{smem}
& &\int {\rm d}^2 {\bf b} ~ e^{-i {\bf Q} \cdot {\bf b} } \int {\rm d} 
{\bf r}_{\rm I}~{\rm d}{\bf r}_{\rm II} \,\Psi^*_{\rm I,f}({\bf r}_{\rm I})\, 
\Psi^*_{\rm II,f} ({\bf r}_{\rm II}) V_{\rm T}({\bf r}_{ik{\rm T}}) 
V_{\rm T}({\bf r}_{jl{\rm T}})\,\Psi_{\rm I,i}({\bf r}_{\rm I}) 
\Psi_{\rm II,i}({\bf r}_{\rm II}) \nonumber\\
& &= {1 \over {(2\pi)^2}} \int {\rm d}^2 {\bf k}_{\rm T} {\widetilde V}_{\rm T} 
({\bf k}_{\rm T}) {\widetilde V}_{\rm T}({\bf k}_{\rm T}-{\bf Q}) F^{\rm M}_{ij}
({\bf Q}, {\bf k}_{\rm T}) 
F^{\rm M}_{kl}({\bf Q}, {\bf k}_{\rm T})\,,~ (i,j=1,2;\,k,l=3,4)\,,
\end{eqnarray}
where the superscript in $F^{\rm M}_{kl}({\bf Q})$ labels the
colliding hadron as a meson M or proton P,
and
\begin{eqnarray}
\label{vffm1}
F^{\rm M}_{ij}({\bf Q}, {\bf k}_{\rm T}) = \sum_{n,n'} a_n a_{n'} \bigg( 
{ {2\sqrt{nn'}} \over {n+n'}}\bigg)^{3/2} \exp\bigg\{-{{[\,{\bf k}_{\rm T}+
C_{{\rm I}i} C_{{\rm I}j}({\bf Q} - {\bf k}_{\rm T})]^2} \over {8(n+n')\,
\beta_{\rm I}^2}} \,\bigg\}\,,
\end{eqnarray}
\begin{eqnarray}
\label{vffm2}
F^{\rm M}_{kl}({\bf Q}, {\bf k}_{\rm T}) = \sum_{m,m'} a_m a_{m'} \bigg( 
{ {2\sqrt{mm'}} \over {m+m'}}\bigg)^{3/2} \exp\bigg\{-{{[\,{\bf k}_{\rm T}+
C_{{\rm II}k} C_{{\rm II}l}({\bf Q} - {\bf k}_{\rm T})]^2} \over {8(m+m')\,
\beta_{\rm II}^2}} \,\bigg\}\,.
\end{eqnarray}

\subsection{The meson$-$proton and proton$-$proton case}

The three-dimensional coordinate ${\bf r}_k$ of a quark in a proton
can be written in terms of Jacobi coordinates $\eeta=(\eeta_{\rm T},
\eta_z)$ and $\bxi=(\bxi_{\rm T},\xi_z)$ \cite{Dol92}.  Its transverse
component is given by
\begin{equation}
\label{coorp}
{\bf r}_{k{\rm T}} = - { {\bf b} \over 2} + { {\delta_{{\rm II}k}} \over 6} 
{\bf \eeta}_{\rm T} + { {C_{{\rm II}k}} \over 2} {\bf \bxi}_{\rm T} \,,
\end{equation}
where
\begin{eqnarray}
\label{coorpc}
\delta_{{\rm II}k} = 
\begin{cases}
2, & ~~~~~k = 3 , \\
-1, & ~~~~~k = 4,\, 5 ,
\end{cases}
~~~~~~~~~~ C_{{\rm II}k} = 
\begin{cases}
0, & ~~~~~k = 3 , \\
1, & ~~~~~k = 4 , \\ 
-1, & ~~~~~k = 5 \,.
\end{cases}
\end{eqnarray}
The relative transverse coordinate of the quark (or antiquark) $i$ in 
meson I and the quark $k$ in proton II is
\begin{eqnarray}
\label{rcoorp}
& &{\bf r}_{ik \rm T}={\bf r}_{i \rm T} - {\bf r}_{k \rm T} = {\bf b} +
{C_{{\rm I}i} \over 2} {\bf r}_{\rm IT} - { {\delta_{{\rm II}k}} \over 6}
{\bf \eeta}_{\rm T} - { {C_{{\rm II}k}} \over 2} {\bf \bxi}_{\rm T} \,.
\end{eqnarray}
We use a simple Gaussian wave function for a proton.  In terms of
Jacobi coordinates, it can be expressed as \cite{Dol92}
\begin{equation}
\label{wfp}
\Psi ({\eeta}, {\bxi})=(\,24\sqrt{3}\,\pi^3\langle\,r_p^2\,\rangle^3\,
)^{-1/2} \exp[(-\eeta^2/12 -\bxi^2/4)/\langle\,r_p^2\,\rangle\,] \,.
\end{equation}
From Eqs.  (\ref{ampl}), (\ref{ftrans2}), (\ref{cssp}), (\ref{wfm}),
and (\ref{coorpc}) --- (\ref{wfp}), the spatial matrix element for a
meson-proton scattering is
\begin{eqnarray}
\label{smemp}
\int {\rm d}^2 {\bf b} ~ e^{-i {\bf Q} \cdot {\bf b} } &\int& {\rm d}
{\bf r}_{\rm I}~{\rm d}\eeta~{\rm d}\bxi \,\Psi^*_{\rm I,f}({\bf r}_{\rm I})\,
\Psi^*_{\rm II,f} (\eeta, \bxi) V_{\rm T}({\bf r}_{ik{\rm T}})
V_{\rm T}({\bf r}_{jl{\rm T}})\,\Psi_{\rm I,i}({\bf r}_{\rm I})
\Psi_{\rm II,i}(\eeta, \bxi) \nonumber\\
& = &{ 1 \over {(2\pi)^2}} \int {\rm d}^2 {\bf k}_{\rm T} {\widetilde V}_{\rm T} 
({\bf k}_{\rm T}) {\widetilde V}_{\rm T}({\bf k}_{\rm T} - {\bf Q}) F^{\rm M}_{ij}
({\bf Q}, {\bf k}_{\rm T})
F^{\rm P}_{kl}({\bf Q}, {\bf k}_{\rm T})\,, \\[2ex]
& &( i,j = 1,2;\, k,l=3,4,5)\,,\nonumber
\end{eqnarray}
where
\begin{eqnarray}
\label{vffp}
F^{\rm P}_{kl}({\bf Q}, {\bf k}_{\rm T}) = 
\begin{cases}
\exp \big ( -{{\langle\,r_p^2\,\rangle}\over 6} {\bf Q}^2 \big ), 
& ~~~~~~k = l , \\
\exp \big \{ -{{\langle\,r_p^2\,\rangle}\over 6} \big [{\bf Q}^2 - {3 
\over 2}{\bf Q}\cdot ({\bf 2k}_{\rm T}) + {3 \over 4} (2 {\bf k}_{\rm 
T})^2 \big ] \big \}, & ~~~~~~k \ne l \,.
\end{cases}
\end{eqnarray}
For a proton-proton scattering, the spatial matrix element is 
\begin{eqnarray}
\label{smepp}
& &\int {\rm d}^2 {\bf b} ~ e^{-i {\bf Q} \cdot {\bf b} } \int {\rm d} 
\eeta_{\rm I}~{\rm d}\bxi_{\rm I}\,{\rm d}\eeta_{\rm II}~{\rm d}\bxi_{\rm 
II} \,\Psi^*_{\rm I,f}(\eeta_{\rm I}, \bxi_{\rm I})\, \Psi^*_{\rm II,f} 
(\eeta_{\rm II}, \bxi_{\rm II}) V_{\rm T}({\bf r}_{ik{\rm T}}) 
V_{\rm T}({\bf r}_{jl{\rm T}})
\,\Psi_{\rm I,i}(\eeta_{\rm I}, \bxi_{\rm I}) \Psi_{\rm II,i}(\eeta_{\rm II}, 
\bxi_{\rm II}) \nonumber\\
& &= {1 \over {(2\pi)^2} } \int {\rm d}^2 {\bf k}_{\rm T} {\widetilde V}_{\rm T} 
({\bf k}_{\rm T}) {\widetilde V}_{\rm T} ({\bf k}_{\rm T} - {\bf Q}_{\rm T}) 
F^{\rm P}_{ij}({\bf Q}, {\bf k}_{\rm T}) \,F^{\rm P}_{kl}({\bf Q}, 
{\bf k}_{\rm T})\,,~~~( i,j = 1,2,3;\, k,l=4,5,6)\,.
\end{eqnarray}

\section{Total Hadron-Hadron Cross Sections}

After we have evaluated the color matrix element, the spin matrix
element, and the spatial matrix element, we can now determined the total
hadron-hadron cross section as
\begin{eqnarray}
\label{totcs}
\sigma_{\rm tot} = { 1 \over 16} \sum_{i,j,k,l} C_{f,ijkl} \,I_{ijkl} \,,
\end{eqnarray}
where
\begin{eqnarray}
\label{colf}
C_{f,ijkl} = C_{ijkl} \sum_{a,b=1}^8 \langle\,C_{\rm I}~|\,\lambda^a_i 
\lambda^b_j \,|~C_{\rm I}\,\rangle \langle\,C_{\rm II}~|\,\lambda^a_k 
\lambda^b_l\,|~C_{\rm II}\,\rangle \,,
\end{eqnarray}
\begin{eqnarray}
\label{ijkl}
I_{ijkl} &=& { 1 \over {(2\pi)^2}} \int {\rm d}^2 {\bf k}_{\rm T} 
{\widetilde V}_{\rm T} ({\bf k}_{\rm T}) {\widetilde V}_{\rm T} ({\bf k}_{\rm T}) 
F_{ij}^{\rm I}({\bf Q}=0, {\bf k}_{\rm T}) 
F_{kl}^{\rm II}({\bf Q}=0, {\bf k}_{\rm T}) 
\nonumber\\
&=& { 1 \over {(2\pi)^2}} \int {\rm d}^2 {\bf k}_{\rm T} \Bigg \{ \bigg [ 
{ {4\pi \alpha_s} \over { {\bf k}^2_{\rm T} + \mu^2 }} + {{6\pi b_0} 
\over {({\bf k}^2_{\rm T} + \mu^2 )^2 } } \bigg]^2 
\nonumber\\
& & + \langle\,s^{(2)}_{ikjl}\rangle  
{{(8\pi \alpha_s)^2} \over {9 m_i m_k m_j m_l} }\,
e^{-{\bf k}^2_{\rm T} /2 d^2} \Bigg\} 
F_{ij}^{\rm I}({\bf Q}=0,{\bf k}_{\rm T}) 
F_{kl}^{\rm II}({\bf Q}=0,{\bf k}_{\rm T}) \,.
\end{eqnarray}
The superscript $\rm I$ (or $\rm II$) is either $\rm M$ (meson) or
$\rm P$ (proton), and $\langle\,s^{(2)}_{ikjl}\,\rangle$ is the
second-order spin-spin matrix element averaged over all polarization
states of hadrons I and II,
\begin{eqnarray}
\label{a2smf}
\langle\,s^{(2)}_{ikjl}\,\rangle = {1 \over N_{\chi} } \sum_{\chi_{_{\rm I, II}}}
\langle\,\chi_{_{\rm I}}\,\chi_{_{\rm II}}~|\,({\bf s}_i \cdot {\bf s}_k)(
{\bf s}_j \cdot {\bf s}_l) \,|~\chi_{_{\rm I}}\,\chi_{_{\rm II}}\,\rangle \, .
\end{eqnarray}
The polarization average of the first-order spin-spin matrix elements
$\langle\,s^{(1)}_{ik}\, \rangle$ is zero.

If we assume an unscreened linear-confining interaction for
hadron-hadron collision (which corresponds to taking $\mu=0$ in the
$6\pi b_0/({\bf k}_{\rm T}+\mu^2)^2$ term in Eq.\ (\ref{ijkl})), the total
cross section will be singular, since the integral over ${\bf k}_{\rm
T}$ in Eq.\ (\ref{ijkl}) diverges.  From a physical viewpoint, we
expect the interaction between constituents in different hadrons to
occur at relatively large separations.  As a consequence, the
interaction will be subject to screening, due to the production of
virtual $q\bar q$ pairs.  The linear confining interaction for bound
states should thus become a screened confining potential when applied
to the interactions of quarks in a diffractive hadron-hadron
scattering process. Similarly we expect the color-Coulomb interaction
to be screened.  The spatial dependence of the interaction in Eq.\
(\ref{vint}) between quarks and antiquarks of different hadrons should
be different from those inside the same hadron.

A simple way to incorporate screening is to replace ${\bf k}^2$ in the
Fourier transforms of the color-Coulomb and the linear interactions by
${\bf k}^2+\mu^2$, which leads respectively to the Yukawa and
exponential potentials, as given in Eqs.\ (\ref{vint}) and
(\ref{vftrans3}) and the interaction in Eq.\ (\ref{ijkl}) .
Previously, in studying the screened potential and its temperature
dependence inferred from lattice gauge theory, the effective string
tension $b_0 =0.35 $ GeV$^2$ was found to give good descriptions of
meson bound state masses \cite{Won02}.  We shall use this effective
string tension in the present work.

The rms radii of hadrons are model-dependent as the interactions are
sometimes implicitly included in the determination of the effective
rms radius.  We shall use the meson bound state wave functions
obtained in the non-relativistic quark model of Ref.\ \cite{Won01},
where the rms radius is 0.256 fm for a pion and 0.261 fm for a
kaon. (See Table IV of Ref.\ \cite{Won01}, in which the quantity
$\sqrt{\langle r^2 \rangle}$ in the fourth column is the rms $q$-$\bar
q$ separation, which is twice the rms meson radius.)  In our present
model in which a hadron-hadron reaction takes place via a finite-range
interaction between constituents in one hadron with constituents of the
other hadron, these hadron constituent radii can be considered ``core''
radii.  These pion and kaon rms ``core'' radii are smaller than the
corresponding values of 0.61 fm and 0.54 fm found in the geometrical
model of Chou and Yang \cite{Cho68,Cho79} as well as the values of
0.56 fm and 0.53 fm determined from electromagnetic measurements
\cite{Dal77,Dal80}.  These differences arise from the fact that Chou
and Yang describe hadrons as a geometrical droplet with an essentially
local zero-range interaction between a constituent of one hadron with
the constituent of the other hadron.  In the electromagnetic
measurement, the photon also fluctuates into a $\rho$ meson, and the
larger electromagnetic rms radii include the additional effects of
this fluctuation.

It is reasonable to assume that the hadron radius $r_{\rm hadron}({\rm
EM})$, obtained in an electromagnetic measurement, is the sum of the
hadron radius used here (the core radius) and an effective rho meson
radius (which is of order $1/m_\rho$),
\begin{eqnarray}
r_{\rm hadron}({\rm EM})=({\rm hadron~core~radius}) 
+ ({\rm effective~}\rho{\rm~meson~radius}).
\end{eqnarray}
Based on this assumption, the difference of of the proton and pion rms
electromagnetic radii is equal to the difference of their
corresponding rms (core) radii,
\begin{eqnarray}
 \sqrt{\langle r_p ^2 ({\rm EM}) \rangle } 
-\sqrt{\langle r_\pi^2 ({\rm EM}) \rangle } 
= \sqrt{\langle r_p ^2 \rangle }
 -\sqrt{\langle r_\pi^2 \rangle }~.
\end{eqnarray}
The proton and pion rms radius obtained in electromagnetic
measurements are 0.81 fm \cite{Wal94} and 0.56 fm \cite{Dal77}
respectively.  Using the meson bound states wave functions in Ref.\
\cite{Won01} for which $\sqrt{\langle r_\pi^2 \rangle }=0.256$ fm, the
above relation gives a proton rms radius of $\sqrt{\langle r_p^2
\rangle }=0.51$ fm which we shall adopt in our present analysis.  This
value of the proton rms radius is close to the value of $\sqrt{\langle
r_p^2 \rangle }=0.48$ fm found by Copley, Karl, and Obryk
\cite{Cop69}, and Koniuk and Isgur \cite{Kon80} in their quark-model
description of baryon electromagnetic transition amplitudes.  We shall
use the quark masses and the hyperfine smearing parameter $d$ that are
employed in the meson bound state wave function calculations in
Ref.\cite{Won01}: $m_u=m_d=0.334$ GeV, $m_s=0.575$ GeV, $m_c=1.776$
GeV, $m_b=5.102$ GeV, and $d=0.897$ GeV.

We search for the screening mass $\mu$ and the strong coupling
constant parameter $\alpha_s$ by fitting the experimental cross
sections of $\sigma_{\rm tot}(p+p)=39.0$ mb, $\sigma_{\rm
tot}(\pi+p)=24.0$ mb, and $\sigma_{\rm tot}(K+p)=20.5$ mb at $\sqrt{s}
=20$ GeV \cite{Pdg02}.  We obtain $\mu=0.425$ GeV, and
$\alpha_s=0.495$ by using the Levenberg-Marquardt method \cite{Pre92}.

The screening mass $\mu=0.425$ GeV falls in between $m_\pi$ and
$m_\rho$ studied by Dolej\v si and H\"ufner \cite{Dol92}.  It is
interesting to note from Fig.\ 3 of their work that a screening mass
between $m_\pi$ and $m_\rho$ gives the best correlation between the
total cross section and the effective elastic scattering slope
parameter ${\rm b}_{\rm eff}$.  It is therefore important to extend
the present calculations to the analysis of elastic scattering in
future work.

The screening length $1/\mu=0.464$ fm is the same physical quantity as
the correlation length $a$ studied by Landshoff and Nachtmann
\cite{Lan87}.  Using a single gluon propagator with a screening mass
$\mu$, they noted that $1/\mu$ must be much smaller than the radius of
a hadron, for the additive quark model to be valid.  Our numerical
analysis, with the full non-perturbative interaction of Eq.\
(\ref{vint}), leads to a range $1/\mu$ that is not small compared to
the radius of a hadron; we should not expect our results to coincide
completely with those of the additive quark model in all aspects.  

The value of $\alpha_s=0.495$ is slightly smaller but close to the
conventional strong interaction coupling constant of $\alpha_s=0.6$
used in non-relativistic quark models for light quarks \cite {Bar92}.

We shall first focus our attention on the total hadron-hadron cross
sections at $\sqrt{s}= 20$ GeV in Fig.\ 2 and then discuss the energy
dependence in Eqs.\ (\ref{highe1})-(\ref{highe3}) and Fig.\ 3.  The
experimental $pp$, $\pi p$, and $K p$ total cross sections at
$\sqrt{s}=20$ GeV can be roughly represented by $\sigma_{\rm tot}\sim
5.86 (\sqrt{\langle r_{\rm I}^2\rangle} +\sqrt{\langle r_{\rm
II}^2\rangle})^2$.  While the theoretical total cross sections
obtained in the present TGMP calculations tend to increase with the
square of the sum of the rms radii, there are however variations that
arise from the spin-spin interaction.  The latter quantity depends on
the masses of the interacting quarks and the strength of the spin-spin
interaction involving strange quarks is smaller than that of light
quarks; consequently, $\sigma_{\rm tot} (\pi+p) > \sigma_{\rm tot}
(K+p)$, as shown in Fig.\ 2.

Our total cross section for $\rho+p$ at $\sqrt{s}=20$ GeV is 27.0
mb, which is slightly larger than the total $\pi + p$ cross section of
24.5 mb.  In the TGMP, the spin-spin interaction affects the $\rho$
meson cross sections in two ways.  First, as the quark-antiquark
potential is weaker for the $\rho$ meson than for the pion due to the
spin-spin interaction, the $\rho$ rms radius is 0.385 fm, which is
substantially larger than the pion rms radius of 0.256 fm.  Secondly,
the spin-spin interaction also affects the interaction strength
between a quark or antiquark in the $\rho$ meson and a constituent in
the other hadron.  Hence, the TGMP predicts $\sigma_{\rm tot}
(\rho+p)$ slightly greater than $ \sigma_{\rm tot} (\pi+p)$.  This
result is slightly greater than the prediction of the additive quark
model which gives
\begin{eqnarray}
\sigma_{\rm tot} (\rho+p)({\rm additive~quark~model})
={1 \over 2} [\sigma_{\rm tot}(\pi^+ + p)+\sigma_{\rm tot}(\pi^- + p)]
\sim 24.2 {\rm ~~~mb}.
\end{eqnarray}

It is interesting to note that our TGMP total cross sections for
$\pi$+$\pi$, and $\rho$+$\pi$ are respectively 27.4 and 20.8 mb,
which are significantly larger than the predictions of of about
$2\sigma_{\rm tot}(\pi+N)/3\sim 16$ mb from the additive quark model.
The large magnitude of the total cross sections in the present
calculation arises from the spin-spin interaction, as the cross
sections become much smaller if the spin-spin interaction is turned
off.

Experimental information of the $\rho$+$p$ cross section can be inferred
from $\rho$-meson photoproduction cross sections.  The extracted total
cross section depends on the coupling constant $f_\rho^2/4\pi$ and the
value of $\eta_\rho$, which is the ratio of the real and imaginary
parts of the scattering amplitude \cite{Bau78}.  It also depends on
the generalizations of the Vector Dominance Model in which one may
include higher resonances and off-diagonal matrix elements
\cite{Pau98,Bia99}.  

Using the simple VDM with $\eta_\rho$ assumed to be zero and
$f_\rho^2/4\pi$ set to be 2.20, the ZEUS Collaboration found
\begin{eqnarray}
\sigma_{\rm tot}(\rho_0 +p)=28.0 \pm 1.2 {\rm ~mb}
\end{eqnarray}
for $\sqrt{s}= 70 $ GeV \cite{Der95} which is close to the value of
$\sigma_{\rm tot}(\rho +p)=30.5 {\rm ~mb}$ obtained in the present
calculation for $\sqrt{s}= 70 $ GeV (see Fig.\ 3).

At low energies, the experimental data cannot be compared directly
with the present theoretical prediction. One can nevertheless get some
idea on the magnitude of $\sigma_{\rm tot}(\rho+p)$. Using
photoproduction data from a number of different nuclei, McClellan
$et~al.$ \cite{McC69} found earlier
\begin{eqnarray}
\sigma_{\rm tot}(\rho_0 +p)=38.0 \pm 3 {\rm ~mb}
\end{eqnarray}
and $f_\rho^2/4\pi=4.40\pm 0.60$ for photons at 6 GeV ($\sqrt{s}=3.48$
GeV).  However, from their subsequent extended measurements and refined
analysis they obtain
\begin{eqnarray}
\sigma_{\rm tot}(\rho_0 +p)=25.9 \pm 1 {\rm ~mb}
\end{eqnarray}
and $f_\rho^2/4\pi=2.52 \pm 0.08$ with $\eta=-0.24$ \cite{Mcc71}
for photons at 4.4-8.8 GeV ($\sqrt{s}=$ 3.02-4.17 GeV).

From an extensive optical model analysis taking the ratio
$\eta_\rho$ to be -0.2, Alvensleben $et~al.$ \cite{Alv69} found
\begin{eqnarray}
\sigma_{\rm tot}(\rho_0 +p)=26.7 \pm 2 {\rm ~mb}
\end{eqnarray}
and $f_\rho^2/4\pi=4.40\pm 0.60$ for photons at 4.6-7.2 GeV
($\sqrt{s}=$ 3.09-3.87 GeV).  Other measurements give
\begin{eqnarray}
\sigma_{\rm tot}(\rho_0 +p)=22.5 \pm 2.7 {\rm ~mb}
\end{eqnarray}
with $f_\rho^2/4\pi=2.52 \pm 0.08$ with $\eta=-0.24$ \cite{Cod75} for
photons at 2.75-4.35 GeV ($\sqrt{s}=2.44-3.01$ GeV).

An independent determination of the $\rho-N$ total cross section,
using the $\gamma+d\to \rho+d$ differential cross sections, was
carried out by Anderson $et~al.$ who obtained
\begin{eqnarray}
\sigma_{\rm tot}(\rho_0 +p)=27.6 \pm 0.6 {\rm ~mb}
\end{eqnarray}
and $f_\rho^2/4\pi=2.80 \pm 0.12$ at a photon energy of 18 GeV
($\sqrt{s}=$5.98 GeV) \cite{And71}.  

In our non-relativistic potential model, the $\rho$ and $\omega$ have
the same spin and spatial wave functions.  Consequently, within the
two-gluon model of the Pomeron, $\sigma_{\rm tot}(\rho+p)=\sigma_{\rm
tot}(\omega+p)$. The value of $\sigma_{\rm tot}(\omega+p)$ extracted
from experimental photoproduction data depends again on the coupling
constant $f_\omega^2/4\pi$ and $\eta_\omega$.  The extracted values of
$\sigma_{\rm tot}(\omega+p)$ range from $33.5\pm5.5$ mb \cite{Beh70}
at 6.8 GeV ($\sqrt{s}=3.694$ GeV) to 25.4$\pm$2.7 mb at 8.3 GeV
\cite{Abr76} ($\sqrt{s}=$4.06 GeV).

Our examination of the experimental $\rho+p$ and $\omega+p$ cross
sections at low energies indicates that they appear to range from 23
mb to 38 mb and the uncertainties are large.  The cross section
depends on energy, and the only data point of $\sigma_{\rm
tot}(\rho+p)$ at $\sqrt{s}=$ 70 GeV that can be direct compared with
the TGMP model indicates approximate agreement between the theoretical
result with experiment.  However, much more work needs to be done to
confront the predictions of the TGMP and the additive quark model with
the results of photoproduction experiments.

For the $\phi+p$ reaction the present results give a theoretical total
cross section of 17.1 mb at $\sqrt{s}=20$ GeV, which agrees with the
naive additive quark model of
\begin{eqnarray}
\sigma_{\rm tot}(\phi+ p)=\sigma_{\rm tot}(K^+ +p) +\sigma_{\rm tot}(K^- +p) 
-\sigma_{\rm tot}(\pi^- +p) \sim 17.0 {\rm ~~~mb}.
\end{eqnarray}
The experimental photoproduction data give a $\sigma_{\rm tot}(\phi
+p)$ which ranges from 9.2 mb to 17.6 mb for photons at 6.4-9.3 GeV
($\sqrt{s}=3.59-4.28$ GeV), depending on the values of $f_\phi^2/4\pi$
and $\eta_\phi$ assumed.

The TGMP gives a theoretical total cross section of $\sigma_{\rm
tot}(J/\psi+p)=$7.6 mb at $\sqrt{s}=$ 20 GeV.  From the absorption of
$J/\psi$ in $p$$A$ collisions, an effective $J/\psi+p$ dissociation
cross section of about 4-6 mb is used to described the dissociation
process \cite{Ger88,Won96,Won98a,Kha96,
Bla96,Cap97,Cas97,Vog98,Nar98,Zha00}.  For the above analysis of
$J/\psi$ production in Pb-Pb collisions at $\sqrt{s}=17.3$ GeV, the
energy at which the $J/\psi$ or its precursor collides with a nucleon
is probably about $\sqrt{s} \sim$8 GeV, corresponding to the collision
of a produced $J/\psi$ nearly at rest with an incident nucleon in the
nucleon-nucleon center-of-mass system. Thus, a theoretical result of
$\sigma_{\rm tot}(J/\psi +p)=$7.6 mb in which the predominant
component may be the dissociation cross section, is approximately
consistent with the experimental dissociation cross section of about
4-6 mb.  It remains necessary to determined more quantitatively how
the $J/\psi+p$ cross section depends on energy, how much of the total
cross section of $J/\psi+p$ can be attributed to dissociation of
$J/\psi$ into an open-charm pair, and whether the observed absorption
of $J/\psi$ inside a nucleus may involve an admixture of charmonium
states with a color-octet component.

The total $J/\psi$+$p$ cross section can also be extracted from the
dipole model of photoproduction \cite{Huf00a}.  In this model, light
cone wave functions for the virtual photon and charmonium are assumed,
and a universal dipole cross section is introduced to fit proton
structure function at small $x$ and a wide range of $Q^2$.  The dipole
model gives $\sigma_{\rm tot}(J/\psi+p)=4.4$ mb at $\sqrt{s}=20$ GeV,
which is lower than the value of 7.6 mb obtained here in the TGMP
model.

Experimental data indicate that at high energies total hadron-hadron
cross sections increase slowly with energy.  This energy dependence is
beyond the scope of the present TGMP model.  We can include such a
dependence phenomenologically, by assuming that the strength
parameters of the interaction depend on energy.  Specifically, we
assume the energy dependence given by Donnachie and Landshoff
\cite{Don92}, and parametrize the energy dependence of the interaction
strengths as
\begin{eqnarray}
\label{highe1}
\alpha_s(\sqrt{s})=\alpha_s(\sqrt{s}=20~{\rm GeV})
f(\sqrt{s}),
\end{eqnarray}
\begin{eqnarray}
\label{highe2}
b_0(\sqrt{s})=b_0(\sqrt{s}=20~{\rm GeV})
f(\sqrt{s}),
\end{eqnarray}
where
\begin{eqnarray}
\label{highe3}
[f(\sqrt{s})]^2=
{ X s^{0.0808} + Y s^{-0.4525} \over
 (X s^{0.0808} + Y s^{-0.4525}) \biggr |_{\sqrt{s}=20~{\rm GeV}} },
\end{eqnarray}
$X=21.70$ mb, $Y=56.08$ mb, and $s$ is in GeV$^2$.  The energy
dependence of our parameters has been so chosen that theoretical
$pp$ cross sections follow the energy dependence of Donnachie and
Landshoff's parametrization of the total $pp$ cross section, which
represents a very good fit to the experimental $pp$ data.

Using such an energy-dependence, we extend our results to energies
near $\sqrt{s}=20$ GeV and beyond.  The lines in Fig. 3 show our
results for total cross sections as a function of $\sqrt{s}$.  The
symbols of circles, open-triangles, solid-triangles, open-squares, and
solid-squares are experimental data for $pp$, $\pi^-p$, $\pi^+p$,
$K^-p$, and $K^+p$ \cite{Pdg02}, respectively.  It can be seen that in
regions where data are available, our results are consistent with
experimental data.  Numerical results for total hadron-hadron cross
sections are given in Table I.

Recently, from the elastic and proton-dissociative $\rho^0$
photoproduction data obtained by the ZEUS Collaboration at HERA
\cite{ZEUS98}, the non-resonant contributions were analyzed in terms
of the fluctuation of a photon into a $\pi^+ \pi^-$ pair and their
interaction with a nucleon \cite{Rys98}.  A model-dependent analysis
of the non-resonant contribution by Ryskin $et~al.$ \cite{Rys98},
based on the Drell-S\"oding approach \cite{Dre60,Sod66}, yields an
effective total $\pi$$p$ cross section of $\sigma(\pi p)= 31 \pm 2
({\rm stat}) \pm 3 ({\rm syst})$ for a pion-proton center-of-mass
energy of about 50 GeV \cite{Arn03}.  The data of the ZEUS
Collaboration is shown as the solid diamond point in Fig.\ 3 and is
consistent with the present results of Fig.\ 3.

\section{Total cross sections for mixed-color charmonium on a pion or a
proton}

A charmonium system that is produced in a parton-parton collision may
be in a nontrivial color state, out of which different pure color
states may be projected \cite{Bod95,Won99}.  This initial state wave
packet may acquire a transverse momentum and will travel through the
nuclear medium. It is therefore clearly of interest to study the
propagation of a colored $c\bar c$ in a color-singlet hadronic medium.

The color wave function of a color-octet meson is
\begin{equation}
\label{coms}
|~C^{8c}\,\rangle={1 \over {\sqrt{2}}} \sum_{\alpha,\,\beta=1}^3 
\lambda^c_{\alpha \beta} |~\alpha ~{\bar \beta}\,\rangle\, , 
~~~~~(c=1,2,\ldots,8) \,.
\end{equation}
For the scattering of a color-octet meson with a color-singlet meson, we
have
\begin{eqnarray}
\label{8+1smm}
\sum_{a,b=1}^8 \langle\, C^{8c}_{\rm I}~|\,\lambda^a_i \lambda^b_j \,
|~C^{8d}_{\rm I} \,\rangle \,\langle\,C^1_{\rm II}~|\,\lambda^a_k 
\lambda^b_l\,|~C^1_{\rm II}\,\rangle = 
\begin{cases}
{32 \over 9} ~\delta^{cd},& ~~~~~~~~~~i = j , \cr
-{4 \over 9} ~\delta^{cd},& ~~~~~~~~~~i \ne j ,\cr
\end{cases}
\end{eqnarray}
and for the scattering of a color-octet meson with a proton we have
\begin{eqnarray}
\label{8+1smp}
\sum_{a,b=1}^8 \langle\, C^{8c}_{\rm I}~|\, \lambda^a_i \lambda^b_j \,
|~C^{8d}_{\rm I} \,\rangle\,\langle\,C^1_{\rm II}~|\,\lambda^a_k 
\lambda^b_l\,|~C^1_{\rm II}\,\rangle = 
\begin{cases}
{32 / 9} ~\delta^{cd},& ~~~~~i = j ,~~ k = l,\cr
-{4 / 9} ~\delta^{cd},& ~~~~~i \ne j ,~~ k =l,\cr
-{16 / 9}~\delta^{cd},& ~~~~~i = j , ~~k \ne l,\cr
{2 / 9}  ~\delta^{cd},& ~~~~~i \ne j, ~~k \ne l\, .\cr
\end{cases}
\end{eqnarray}

The color state of a mixed-color meson can be expressed as
\begin{eqnarray}
\label{csmix}
|~C^{mixed-color}\,\rangle = \gamma_1\,|~C^1\,\rangle + \sum_{c=1}^8 
\gamma_{8c}\,|~C^{8c}\,\rangle \,.
\end{eqnarray}
For a mixed-color meson scattering on a hadron, we have
\begin{eqnarray}
\label{cm8+1}
& &\sum_{a,b=1}^8 \langle\, C^{mixed-color}_{\rm I}~|\,\lambda^a_i 
\lambda^b_j \, |~C^{mixed-color}_{\rm I} \,\rangle\,\langle\,C^1_{\rm II}~
|\,\lambda^a_k \lambda^b_l\, |~C^1_{\rm II}\,\rangle \nonumber\\
& &= \sum_{a,b=1}^8 \bigg [ |\gamma_1|^2 \langle\, C^1_{\rm I}~|\,
\lambda^a_i \lambda^b_j\, |~C^1_{\rm I} \,\rangle + \sum_{c=1}^8 
|\gamma_{8c}|^2 \langle\, C^{8c}_{\rm I}~| \,\lambda^a_i \lambda^b_j\, 
|~ C^{8c}_{\rm I} \,\rangle\, \bigg ] \nonumber\\
& &~~~~\times \langle\,C^1_{\rm II}~|\,\lambda^a_k \lambda^b_l\,|~C^1_{\rm 
II}\,\rangle \nonumber\\
& &= |\gamma_1|^2 \bigg ( \sum_{a,b=1}^8 \langle\,C^1_{\rm I}~|\,
\lambda^a_i \lambda^b_j\, |~C^1_{\rm I}\,\rangle\,\langle\,C^1_{\rm II}~
|\,\lambda^a_k \lambda^b_l\, |~C^1_{\rm II}\,\rangle \bigg ) \nonumber\\ 
& &+ \sum_{c=1}^8 |\gamma_{8c}|^2 \bigg ( \sum_{a,b=1}^8 
\langle\, C^{8c}_{\rm I}~|\,\lambda^a_i \lambda^b_j\, |~C^{8c}_{\rm I}\,
\rangle \,\langle\,C^1_{\rm II}~|\,\lambda^a_k \lambda^b_l\, |~C^1_{\rm 
II}\,\rangle \bigg ) \,.
\end{eqnarray}

From Eqs. (\ref{totcs}) --- (\ref{ijkl}) and (\ref{cm8+1}), the total 
cross section of a mixed-color meson scattering on a hadron can be 
expressed as \cite{Won97}
\begin{eqnarray}
\label{tcs8+1}
\sigma_{\rm tot}=(1-P_8)\,\sigma^{1+1}_{\rm tot}+P_8\,\sigma^{8+1}_{\rm tot}
\, ,
\end{eqnarray}
where $P_8=\sum_{c=1}^8 |\gamma_{8c}|^2$ gives the total amount of
color-octet mixing.

In Fig. 4(a) and 4(b) we show the total scattering cross sections of a
mixed-color charmonium on a pion or on a proton, as a function of the
size of the $c\bar c$ pair and the amount of color mixing at
$\sqrt{s}=20$ GeV.  The wave function of the charmonium state is taken
to be a single Gaussian wave function characterized by a root-mean
square radius.  The cross sections are insensitive to the spin state
of the $c \bar c$ system.  The results in Fig. 4 are the total cross
sections for the collision of a spin-triplet $S=1$ $c\bar c$.  It can be
seen that the total cross section of a color-octet charmonium
($P_8=1.0$) scattering on a hadron remains finite even as the size of
the charmonium approaches zero, and they change only weakly with the
charmonium size.  This is quite different from the case of a
color-singlet charmonium ($P_8=0.0$) scattering from a color-singlet
hadron.  For the case of the scattering of two color-singlet mesons,
if we turn off the spin-spin interaction, the integrand in
Eq. (\ref{ampl}) is proportional to
\begin{eqnarray}
\label{intfsm}
[ V_{\rm T}({\bf r}_{\rm 1T}-{\bf r}_{\rm 3T})
 +V_{\rm T}({\bf r}_{\rm 2T}-{\bf r}_{\rm 4T})
 -V_{\rm T}({\bf r}_{\rm 1T}-{\bf r}_{\rm 4T})
 -V_{\rm T}({\bf r}_{\rm 2T}-{\bf r}_{\rm 3T})
]^2 \,,
\end{eqnarray}
and for the case of the scattering of a color-octet meson with a
color-singlet meson, the integrand is instead proportional to
\begin{eqnarray}
\label{intfcm}
[ V_{\rm T}({\bf r}_{\rm 1T}-{\bf r}_{\rm 3T})
 +V_{\rm T}({\bf r}_{\rm 2T}-{\bf r}_{\rm 4T})
 -V_{\rm T}({\bf r}_{\rm 1T}-{\bf r}_{\rm 4T})
 -V_{\rm T}({\bf r}_{\rm 2T}-{\bf r}_{\rm 3T})
]^2 \nonumber\\
+{9\over 4}[
 V_{\rm T}({\bf r}_{\rm 1T}-{\bf r}_{\rm 3T})
-V_{\rm T}({\bf r}_{\rm 1T}-{\bf r}_{\rm 4T})][
 V_{\rm T}({\bf r}_{\rm 2T}-{\bf r}_{\rm 3T})
-V_{\rm T}({\bf r}_{\rm 2T}-{\bf r}_{\rm 4T})]\,.
\end{eqnarray}
We can see that expression (\ref{intfcm}) does not approach zero when
the size of the color-octet meson approaches zero (${\bf r}_{\rm 1T}
\to {\bf r}_{\rm 2T}$), unlike expression (\ref{intfsm}).  There are
cancellations in the case of color-singlet hadron-hadron scattering,
which are not present in the scattering of a color-octet charmonium
state on color-singlet hadrons, as first pointed out by Dolej\v si and
H\"ufner \cite{Dol92}.  

The color-octet cross sections for $P_8=1.0$ are approximately the
same as those of Dolej\v si and H\"ufner \cite{Dol92}, even though the
potentials used are quite different.  For the collision of the
color-singlet charmonium ($P_8=0.0$) on $\pi$ and $p$, the results at
$\sqrt{\langle r_{c\bar c}^2 \rangle} =0.2$ fm in Fig.\ 4 are close to
the results for $J/\psi+\pi$ and $J/\psi+p$ in Fig.\ 2 as the rms
radius of $J/\psi$ in our calculation is 0.202 fm.

\section{Summary and Conclusions}

We have calculated total hadron-hadron cross sections for the
collisions of $\pi$, $K$, $\rho$, $\phi$, $D$, $J/\psi$, $\psi'$,
$\Upsilon$, $\Upsilon'$, and the proton using the Low-Nussinov model
of two-gluon exchange, as developed by Gunion, Soper, Landshoff,
Nachtmann, Dolej\v si, H\"ufner, and Wong.  A set of meson wave
functions obtained from a non-relativistic quark potential model is
used in our calculations.  A phenomenological potential including
screened color-Coulomb, screened confining, and spin-spin interactions
is employed to describe the gluon exchange between the constituents of
the interacting hadrons. The screening mass and the strong coupling
constant of the potential are obtained by fitting the total cross
sections of $p$+$p$, $\pi$+$p$, and $K$+$p$ collisions at
$\sqrt{s}=20$ GeV.  We extend our results to higher energies using the
phenomenological energy dependence obtained by Donnachie and Landshoff
\cite{Don92}.

We find that the total hadron-hadron cross sections correlate with the
size of the scattering hadron system.  There are, however, important
variations arising from the spin-spin interaction.  This leads to
$\sigma_{\rm tot}(\pi+p)> \sigma_{\rm tot}(K+p)$, a larger $\rho$ rms
radius, and consequently a $\sigma_{\rm tot}(\rho+p)$ that is slightly
greater than $\sigma_{\rm tot}(\pi+p)$.  The experimental total
$\pi+p$ cross section at 50 GeV obtained by the ZEUS Collaboration
agrees with the theoretical result of the TGMP and the experimental
ZEUS data point of the total $\rho+p$ cross section at $\sqrt{s}=70$
GeV is close to the theoretical prediction.

While the TGMP and the additive quark model give about the same cross
sections in some cases, there are significant deviations because the
effective spatial range of the quark-quark interaction in the present
analysis, $1/\mu$, is not small compared to hadron radii.  The total
$\pi$+$\pi$, and $\pi$+$\rho$ cross sections obtained here differ
significantly from those of the additive quark model.  Much more work
remains to be done to confront the predictions of the TGMP and the
additive quark model with experiments.  It will be of great interest
to obtain quantitative experimental measurements of these cross
sections, using for example the production of pions in photon-photon
collisions in which each photon fluctuates into a pair of pions, as in
the Drell \cite{Dre60} and S\"oding \cite{Sod66} approach in photon
reaction processes.

As was previously noted by Dolej\v si and H\" ufner \cite{Dol92}, we
find that the theoretical total cross section for a color-octet meson
is quite large.  The large cross section arises from the fact that in
the interaction of a color-octet meson, the color interactions of both
constituents in the meson interfere constructively, while in a
color-singlet meson, the interactions interfere destructively.
Therefore, a color-octet meson traveling through a color-singlet
nuclear medium is expected to suffer frequent collisions and will lose
a substantial amount of energy.  These results for the color-octet
cross sections may be used to place a constraint on the color-octet
production amplitude of a heavy-quark pair produced in a
nucleon-nucleon collision by studying its propagation in the nuclear
medium.

It will be of interest in future work to study the elastic
differential cross section based on the TGMP, which can be compared
directly with experiment.  It would be useful to develop experimental
and theoretical techniques to infer cross sections of unstable mesons
on various targets, which would be valuable to discriminate between
various theoretical models.

\begin{acknowledgments}
The authors would like to thank Drs.\ T. Barnes, H. Crater, and
C. W. Wong for helpful discussions.  WNZ would also like to thank
Dr. Glenn Young for his kind hospitality at Oak Ridge National
Laboratory.  This research was supported in part by the Division of
Nuclear Physics, U.S. Department of Energy, under Contract
No. DE-AC05-00OR22725, managed by UT-Battelle, LLC, and by the
National Natural Science Foundation of China under Contract
No. 10275015.
\end{acknowledgments}

\begin{table*}
\caption{\label{tab:table1}Numerical results for total hadron-hadron 
cross sections.}

\vspace*{0.2cm}
\begin{ruledtabular}
\begin{tabular}{ccccc}
 &$\sqrt{s}=20$ GeV&$\sqrt{s}=80$ GeV&$\sqrt{s}=140$
 GeV&$\sqrt{s}=200$ GeV\\ &$\sigma_{\rm tot}$ (mb)&$\sigma_{\rm tot}$
 (mb)&$\sigma_{\rm tot}$ (mb)& $\sigma_{\rm tot}$ (mb)\\ \hline
 p$+$p&39.118&45.324&49.089&51.785\\
 $\pi+$p&24.522&28.412&30.772&32.463\\
 K$+$p&19.601&22.711&24.597&25.948\\
 $J/\psi+$p&7.597&8.802&9.533&10.057\\
 $J/\psi+\pi$&3.167&3.669&3.974&4.193\\
 $J/\psi+$K&3.094&3.585&3.883&4.096\\
 $J/\psi+\rho$&4.703&5.449&5.902&6.226\\
 $\pi+\pi$&27.393&31.739&34.375&36.264\\
 $\pi+$K&18.791&21.772&23.581&24.876\\
 K$+$K&13.402&15.528&16.818&17.742\\
 $\rho+$p&26.955&31.231&33.825&35.684\\
 $\rho+\pi$&20.801&24.101&26.103&27.537\\
 $\rho+$K&15.622&18.100&19.604&20.681\\
 $\rho+\rho$&20.397&23.633&25.596&27.002\\
 $\phi+$p&17.104&19.818&21.464&22.643\\
 $\phi+\pi$&9.716&11.257&12.192&12.862\\
 $\phi+K$&8.139&9.430&10.214&10.775\\
 $\phi+\rho$&11.610&13.452&14.569&15.370\\
 $\psi'+$p&15.180&17.588&19.049&20.096\\
 $\psi'+\pi$&5.529&6.406&6.938&7.319\\
 $\psi'+K$&5.577&6.462&6.999&7.383\\
 $\Upsilon+$p&3.546&4.109&4.450&4.694\\
 $\Upsilon+\pi$&1.423&1.649&1.786&1.884\\
 $\Upsilon+K$&1.460&1.692&1.832&1.933\\
 $\Upsilon'+$p&10.903&12.633&13.682&14.434\\
 $\Upsilon'+\pi$&3.895&4.513&4.888&5.156\\
 $\Upsilon'+K$&4.050&4.693&5.082&5.361\\
 $D+$p&18.221&21.112&22.865&24.121\\
 $D+\pi$&13.852&16.050&17.383&18.338\\
 $D+K$&10.540&12.212&13.226&13.953\\
 $D+\rho$&13.435&15.566&16.859&17.786\\
\end{tabular}
\end{ruledtabular}
\end{table*}

\newpage

\begin{figure}
\includegraphics{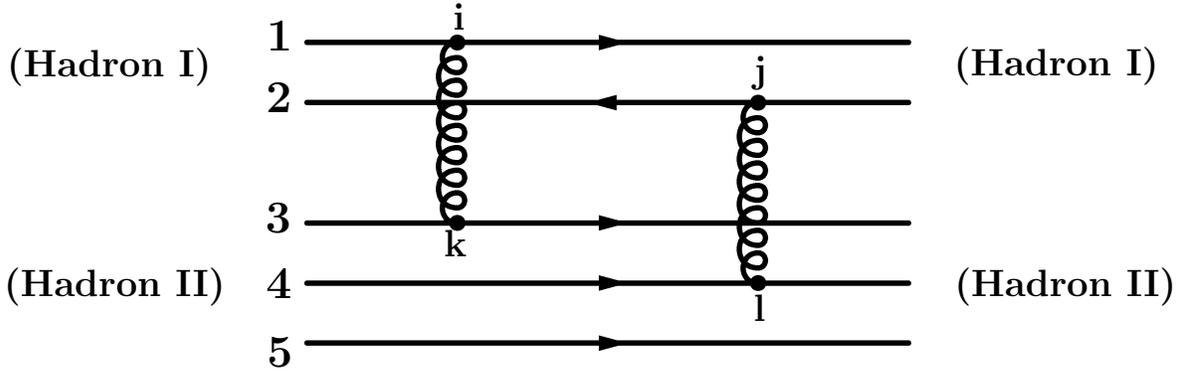}
\caption{\label{fig:fey}  A two-gluon exchange diagram contribution to
the meson-baryon elastic scattering amplitude.}
\end{figure}

\begin{figure}
\includegraphics{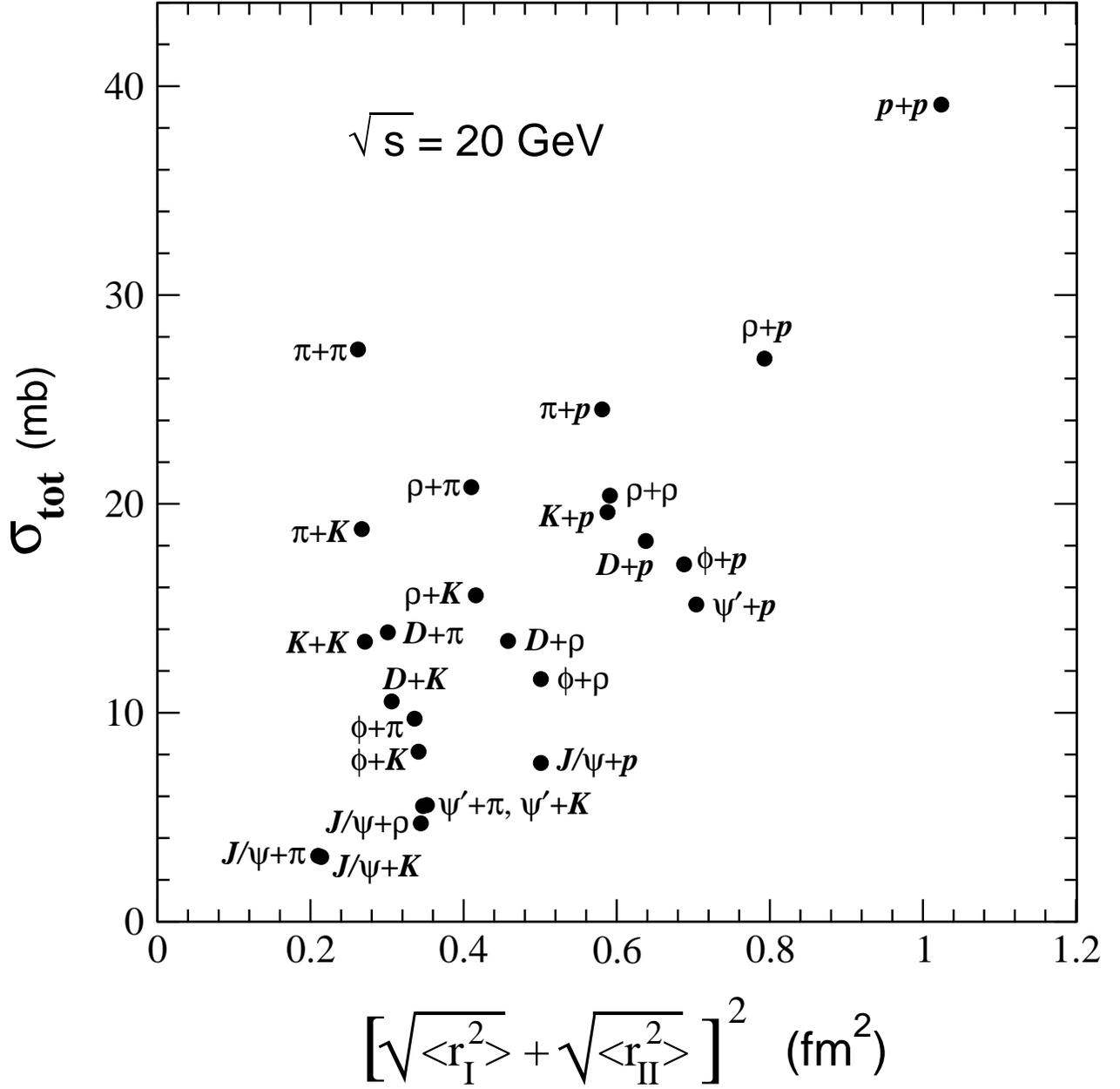}
\caption{\label{fig:xcs} Total hadron-hadron cross sections as a
function of the square of the sum of the rms radii of the colliding
hadrons.  }
\end{figure}

\begin{figure}
\includegraphics{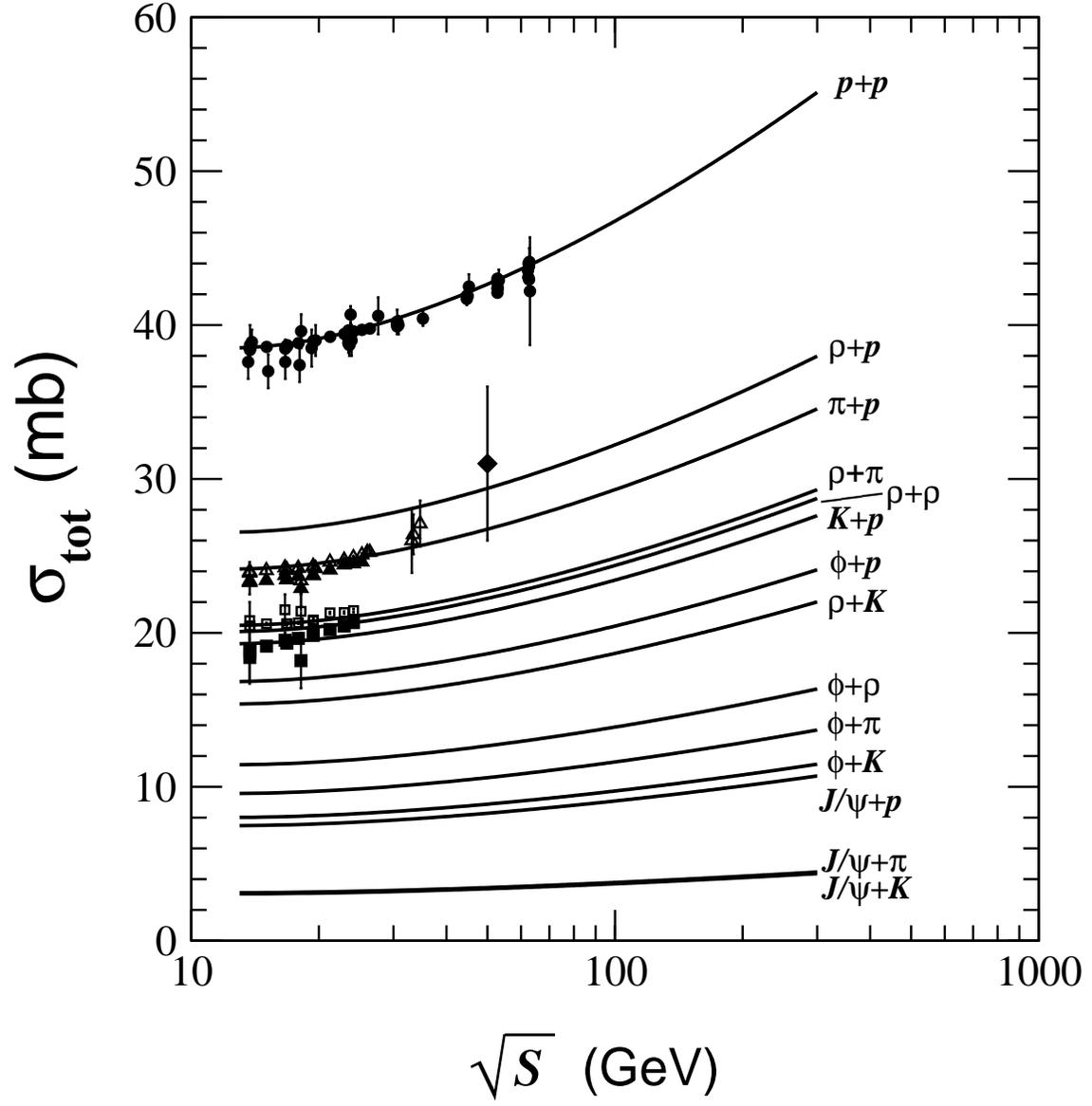}
\caption{\label{fig:xcss} Total hadron-hadron cross sections as a
function of energy. The symbols of circles, open-triangles,
solid-triangles, open-squares, and solid-squares represent
experimental data for $pp$, $\pi^-p$, $\pi^+p$, $K^-p$, and $K^+p$
\cite{Pdg02}.  The solid diamond point is from the $\pi$+$p$ data of
the ZEUS Collaboration \cite{ZEUS98}.}
\end{figure}

\begin{figure}
\includegraphics{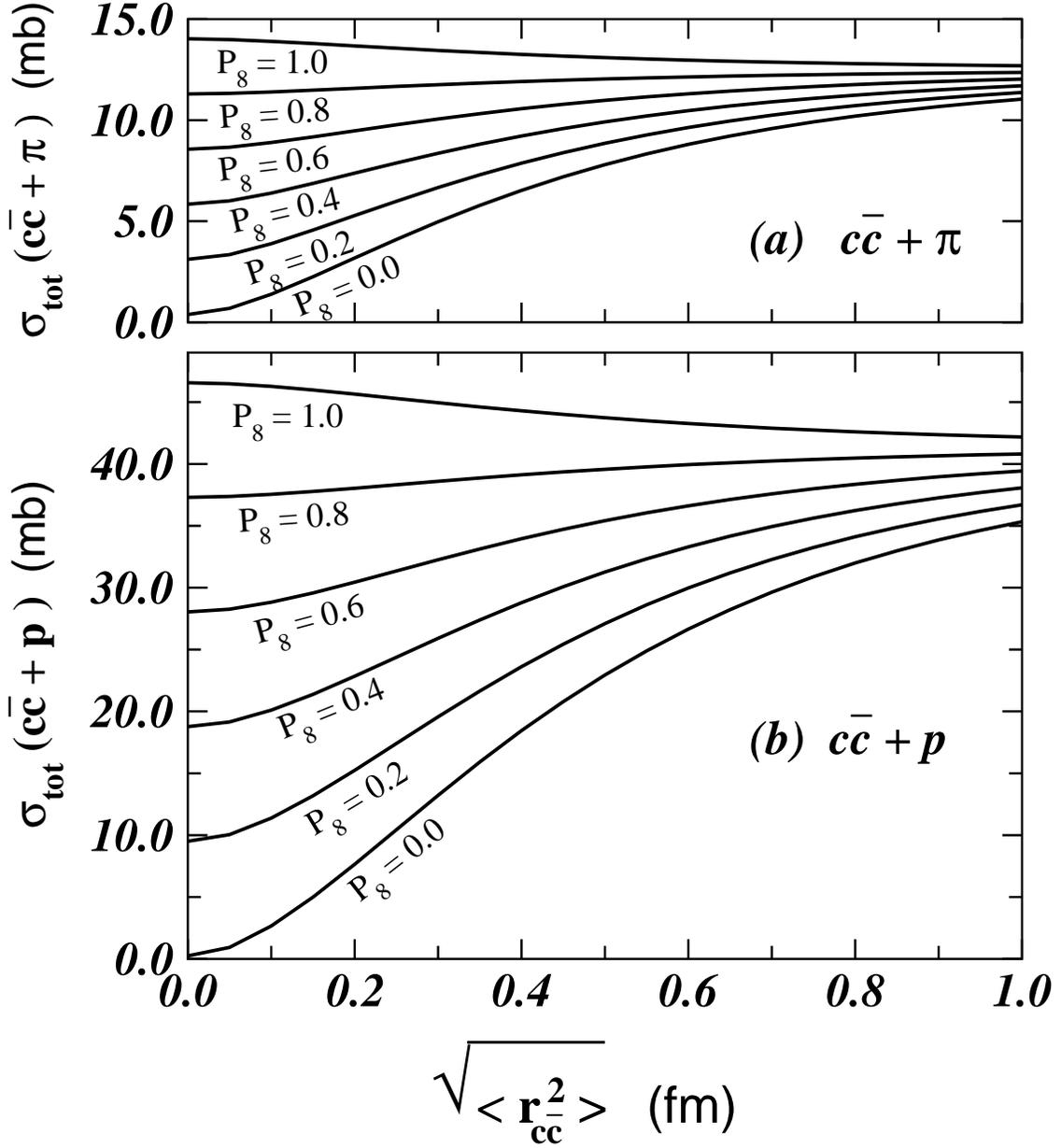}
\caption{\label{fig:cc+pion} (a) The total cross sections for
mixed-color charmonium scattering on a pion as a function of the
charmonium rms radius at $\sqrt{s}=20$ GeV. $P_8$ specifies the
probability of the charmonium color-octet state.  (b) The total cross
sections of mixed-color charmonium scattering on a proton as a
function of the charmonium rms radius at $\sqrt{s}=20$ GeV.}
\end{figure}


\begin{thebibliography}{99}
\bibitem{Arn03} The authors are indebted to Dr. M. Arneodo for
bringing the ZEUS data for $\pi$+$p$ at about 50 GeV \cite{ZEUS98} to
their attention.
\end{thebibliography}
\end{document}